\documentstyle[aps,12pt,epsf]{revtex}
\begin{document}
\begin{center}
{\large\bf Photoproduction and electroproduction of charm at high energies}

\vspace*{1cm}
A.V.Berezhnoy\\
{\normalsize \it Skobeltsin Institute of Nuclear Physics, Moscow State
University, Moscow, Russia}\\
and\\
V.V.Kiselev, A.K.Likhoded\\
{\normalsize \it Russian State Research Center ``Institute of High Energy
Physics'', Protvino, Moscow region, 142284, Russia}\\
{\normalsize \it Fax: 095-2302337, E-mail: kiselev@mx.ihep.su}

\end{center}

\begin{abstract}
We estimate the differential and total cross sections for both the
photoproduction of vector $D^*$-meson and its yield in deep inelastic
scattering at the HERA collider in the framework of model motivated by
perturbative calculations in QCD. The offered model allows us to take into
account higher twists over the transverse momentum of meson at $p_T \sim m_c$
and to correctly approach the dominance of $c$-quark fragmentation at $p_T \gg
m_c$. We consider a possibility for the hadronization of color-octet $c \bar
q$-state into the meson. The combined contribution by the singlet and
octet-color terms results in a good agreement with the experimental data for
both the photoproduction and the production in deep inelastic scattering.
\end{abstract} 

\vspace*{1cm}

\newpage
\section{Introduction.}

New data on the photoproduction and electroproduction of charm in deep
inelastic scattering (DIS) recently obtained by collaborations of ZEUS
\cite{ZEUS} and H1 \cite{H1} at the HERA accelerator, stimulate the discussion
on the following problem: what is an interpretation of these data in the
framework of perturbative QCD (pQCD)? At present the data show the experimental
confirmation of qualitative predictions of pQCD. However, essential
quantitative deviations of measured cross sections from the theoretical
expectations were observed in some kinematical regions, i.e. the discrepancies
in the form of $D^*$-meson spectra were found.

Exploring the perturbative theory in the processes under consideration is
strictly justified due to quite large transverse momenta of $D^*$-mesons,
$p_T\ge m_c$. To the same moment, we cannot completely avoid nonperturbative
terms in the calculations, and the hadronization process of $c$-quarks produced
in hard collisions of initial partons is usually described by means of a
fragmentation function (FF) $D(z,\mu)$, where the parameter $\mu$ determines
the scale of factorization for the probabilities of perturbative subprocess and
binding the quarks into the hadron. The hadronization of quarks is determined
by the dynamics of confinement\footnote{Another region, wherein
phenomenological parameterizations are necessary because of the nonperturbative
character of interactions, is the description of parton distributions in
initial states, though the evolution versus the scale of factorization for the
structure functions allows the consideration in the framework of pQCD.}. The
information on the nonperturbative FF is taken from the data on the production
of charmed particles in $e^+e^-$-annihilation \cite{OPAL,ARGUS}. The dependence
of FF on $\mu$ is determined by the leading logarithmic approximation of QCD
(LO) and corrections to it (NLO).  Calculating the cross sections for the
photoproduction of heavy quarks in the $O(\alpha \alpha_s^2)$-order, some
authors neglect the $D(z,\mu)$ changes versus $\mu \sim p_T$ in the
hadronization \cite{Frixione}. This dependence of FF was taken into account in
other papers \cite{Kniehl}. In this way, the authors used the initial FF at
$\mu=\mu_0\sim$ 1 GeV in the forms of both the ans\" atze based on the
reciprocity relation \cite{Resip} and FF by Peterson et al. \cite{Peterson}.

In pQCD two approaches are used for the calculation of $c\bar c$-pair
production up to the $O(\alpha \alpha_s^2)$-terms. In the first, authors assume
that the light quarks and gluons are the only partons present in the photon and
proton, when the $c$-quarks are produced in the interactions of those partons.
In this way the nonzero mass of $c$-quark is taken into account
\cite{Frixione}.

The second approach suggests that the $c$-quark is the additional ``active''
flavor supposed as massless, and the $c$-quark distributions in the initial
particles are introduced in the form of corresponding structure functions
\cite{Kniehl,Cacciari}.

Note, that in the both ways the hadronization of $c$-quarks is described in the
framework of fragmentation model, which results in the convolution of
perturbative distribution of charmed quarks with the fragmentation function:
\begin{equation}
\frac{d\sigma_{D^*}}{dp_T}=
\int_{2p_T/\sqrt{\hat s}}^1
\left.\frac{d\hat \sigma_{c \bar c}(k_T,\mu)}{dk_T}
\right|_{k_T=\frac{p_T}{z}}\cdot \frac{D_{c \to D^*}(z,\mu)}{z}dz,
\label{fact}
\end{equation}
where $D_{c \to D^*}(z,\mu)$ is the FF normalized by the probability of
$c$-quark transition into the $D^*$-meson $w(c \to D^*)$, which is measured in
the $e^+e^-$-annihilation \cite{OPAL} ($w(c\to D^*)=0.22\pm 0.014\pm 0.014$),
and $\mu$ is the factorization scale for the perturbative partonic cross
section
$d\hat \sigma_{c\bar c}/dk_{T}$.

According to the model of heavy quark fragmentation, the basic condition of its
applicability is $p_T\gg m_c$ or $m_{c\bar c}\gg m_D$ providing a jet-like
shape of measured events. However, the dominant term of statistics is
integrated in the region of $p_T\sim m_c$, where the model of (\ref{fact}) may
not be explored.
  
In the model offered below, we describe the production of $c\bar q$ quark
state in pQCD and use the semilocal duality \cite{Dual} assuming that this
quark state repeats the distributions of hadron with the accuracy up to the
factor depending on the quantum numbers of $c\bar q$ and the hadron. In this
way we consider the complete set of Feynman diagrams corresponding to the
production of $c\bar q$-system in the fixed order over $\alpha$ and $\alpha_s$
(so the leading order is the fourth one). Then we quite successfully describe
the production of $D^*$-mesons at $p_T\sim m_c$ as well as reliably and
correctly approach the results of fragmentation model at $p_T\gg m_c$. Such the
agreement of model estimates with the experimental data can be reached, if we
add the term of octet-color state for the production of $c\bar q$ pair  along
with the contribution of the singlet-color state.

In the framework of pQCD taking into account the full $Q^2$-dependence of the
matrix element, exact calculations give a good description of data on the DIS
production of charm. This estimate is carried out up to NLO in refs.
\cite{qcddis}. Despite of the qualitative success in such the approach,
quantitative deviations between the predictions and measured differential cross
sections take place as in the photoproduction of charm. In present paper we
calculate the differential cross sections for the $D^*$ production in DIS in
the framework of model taking into account the mentioned contribution of octet
operator into the production of charmed quark-light quark pair.

\section{Fragmentation function.}

Among the well-known parameterizations of FF, the model by Peterson et al.
\cite{Peterson} argues the following form:
\begin{equation}
D(z)=N\frac{1}{z(1-\frac{1}{z}-\frac{\epsilon}{(1-z)})^2},
\label{Peter}
\end{equation}
where $N$ is the normalization factor, and $\epsilon$ is the free
phenomenological parameter dependent of $\mu$.
This FF quite accurately describes the data on the production of $B$- and
$D$-mesons in $e^+e^-$-annihilation at high energies, where the fragmentation
mechanism dominates \cite{Frag}.

According to motivations by the authors of parameterization (\ref{Peter}), the
behaviour versus $z$ in Eq.(\ref{Peter}) is determined by the virtuality in the
propagator of heavy $c$-quark. Indeed, let $P_{D^*}$ and $P_{jet}$ denote the
momenta of $D^*$-meson and an associated jet, respectively. Then the
denominator of perturbative propagator has the form
$$
m_c^2-(P_{D^*}+P_{jet})^2.
$$
Expanding this expression in small parameters $\frac{m_{D^*}}{E_{D^*}}$
and $\frac{m_{jet}}{E_{jet}}$ and defining $z=\frac{E_{D^*}}{E_{D^*}+E_{jet}}$,
we get
$$
   m_c^2-(P_{D^*}+P_{jet})^2\approx
   m_c^2-\frac{m_{D^*}^2}{z}-\frac{m_{jet}^2}{1-z}\approx
   m_c^2-\frac{m_c^2}{z}-\frac{m_{jet}^2}{1-z}\sim
   1-\frac{1}{z}-\frac{m_{jet}^2}{m_c^2}\cdot \frac{1}{1-z}.
$$
In the parameterization of (\ref{Peter}) the quantities $N$ and $\epsilon$ are
treated as phenomenological and nonperturbative parameters, though we can clear
see that $\epsilon$ corresponds to the mass ratio squared for the jet mass over
the quark mass, whereas the jet mass logarithmically grows with the increase of
total energy in the $e^+e^-$-annihilation.

The consistent consideration of heavy quark fragmentation \cite{Frag} with
account for both the quantum numbers of meson in the final state and the
QCD-structure of vertex on Fig.~1, leads to the following analytical
expression\footnote{The quark-meson vertex contains the form factor which makes
the loop calculations free off any divergences: ultraviolet or infrared. So,
the calculations are free off any regularization procedure, and we do not need
any discussion on the regularization. This is very well known from
ref.\cite{Frag}. To the leading approximation the quarks entering the meson
have the same velocities and are on mass-shells (see ref.\cite{Frag}). This
takes place if we calculate the fast or valence components of the meson. The
perturbation theory is useless, if one calculates the soft quark-gluon sea,
which is out off current study, since we restrict the gluon virtualities and
consider the fast degrees of freedom in the meson. According to the
factorization theorem in perturbative QCD at high $p_T$ the restricted number
of diagrams contributes. It is well known that the only hard diagram (Fig.~1)
contributes to the fragmentation in the special axial gauge (see
ref.\cite{Frag}). In the covariant gauge two hard diagrams do. The calculations
of fragmentation functions were done in both gauges, and the FF does not depend
on the gauge.}:
\begin{eqnarray}
\displaystyle
D_{\bar c \to D^*}(z)&=&
\frac{8\alpha_s^2\langle O_{(\ref{fact})} \rangle}{27 m_q^3}
\frac{rz(1-z)^2}{(1-(1-r)z)^6}
[2-2(3-2r)z+3(3-2r+4r^2)z^2-
\nonumber
\\
&&
2(1-r)(4-r+2r^2)z^3+
(1-r)^2(3-2r+2r^2)z^4],
\end{eqnarray}
where  $r=m_q/(m_q+m_c)$ and
$$
\langle O_{(\ref{fact})} \rangle =
\frac{1}{12 M} \left(-g^{\mu\nu}+\frac {p^\mu p^\nu}{M^2}\right)\;
\langle D^*(p)|(\bar c \gamma_\mu q) (\bar q \gamma_\nu
c)|D^*(p)\rangle,
$$
which corresponds to the square of wave function at the origin for two quarks
in the framework of nonrelativistic potential model: $\langle O_{(\ref{fact})}
\rangle|_{NR} =|\Psi(0)|^2$. The effective mass of light quark $m_q$ determines
the form as well as the normalization of FF:
\begin{equation}
w(c \to D^*)=\int_0^1 D_{c\to D^*}(z)dz=
\frac{\alpha_s^2(\mu_R)\langle O_{(\ref{fact})}(\mu_R)\rangle
}{m_q^3} \cdot I(r),
\label{prob}
\end{equation}
where
\begin{eqnarray}
I(r) &=& \frac{8}{27}\left [
\frac{24+109r-126r^2-174r^3-89r^4}{15(1-r)^5}+\right. \nonumber \\
&+& \left. \frac{r(7-4r-3r^2+10r^3+2r^4)}{(1-r)^6}\ln(r)\right ].
\end{eqnarray}
In the model under consideration we put the probability of fragmentation
$w(c\to D^*)$ independent of the scale $\mu_R$, so that $w$ remaines the
constant value extracted from the experimental data mentioned.

We argue the following motivation: The distributions of partons inside the
heavy-light meson have distinct terms corresponding to the valence quarks. So,
for the neutral charmed meson $c\bar u$, say, in the infinite momentum frame
these contributions are given by the functions depending on the fractions of
longitudinal momentum and the transverse momenta of partons. The valence terms
are determined by
\begin{eqnarray}
p_c^v(x,p_\perp) &=& p_c(x,p_\perp)-p_{\bar c}(x,p_\perp),\\
p_{\bar u}^v(x,p_\perp) &=& p_{\bar u}(x,p_\perp)-p_{u}(x,p_\perp),
\end{eqnarray}
so that the functions yield the average fractions 
\begin{eqnarray}
\langle x^v_c\rangle =\int d^2 p_\perp dx\; x\cdot p_c^v(x,p_\perp) &\approx &
\frac{m_c}{M_D},\\
\langle x^v_{\bar u}\rangle =\int d^2 p_\perp dx\; x\cdot p_{\bar
u}^v(x,p_\perp) &\approx & \frac{\bar \Lambda}{M_D},
\end{eqnarray}
with $\langle x^v_c\rangle+\langle x^v_{\bar u}\rangle \approx 1$, and $\bar
\Lambda$ denotes the binding energy of heavy quark inside the meson. Then, in
the model we neglect the dispersions of quark fractions and put the valence
quark velocities equal to each other: $v_c = v_{\bar u}$. Generally, we
can write down the following representation for the Fock state of charmed
meson:
\begin{equation}
|D\rangle = W_{(\ref{fact})} |(c\bar u)_{(\ref{fact})}\rangle + W_{(8)} |(c\bar
u)_{(8)}\rangle ,
\end{equation}
where the quark pairs $(c\bar u)$ carry the mentioned fractions of meson
momentum, and they belong to two possible color states: the singlet or octet,
and the operators $W_{(1,8)}$ generate the quark-gluon sea inside the meson at
the characteristic scale of bound state $\mu_R$. There are several qualitative
features of this representation. First, the operators $W$ are nonperturbative.
Therefore the scale of interactions in $W$ is about the scale of confinement
$\Lambda_{QCD}$. In contrast to the heavy-light meson, for the heavy quarkonium
there are the counting rules due to the small relative velocity of heavy
quarks, i.e. due to the smallness of typical inverse size of bound state with
respect to the quark mass. In the heavy quarkonium the $W_{(8)}$ contribution
is suppressed by a power of relative velocity \cite{OCT}. In the heavy meson
under consideration, the octet contribution is not suppressed, and it is of the
same order as the singlet one. Second feature concerns for the excitations in
the system of valence quarks. For the heavy quarkonium the colored $P$-wave
states contribute, since their sizes are still small enough to propagate with
no decay into the pairs of heavy mesons. In the heavy-light system the $P$-wave
states are suppressed because they are heavy and have large sizes leading
to an instability. Then, for the production of charmed mesons the contribution
of singlet and octet $S$-wave state is essential only. The operators
$W_{(1,8)}$ are infrared and cannot be calculated in perturbative QCD. However,
they do not influence onto the distributions of meson at the transverse momenta
higher that the charmed quark mass, $p_T\sim m_c$. To get rid of the
nonperturbative quark-gluon sea from the calculations, we have to restrict the
gluon virtualities $k_g^2> \mu^2_R$, so that $\Lambda_{QCD}^2/\mu_R^2\ll 1$. We
suppose $\mu_R \sim m_\rho$, and consider the massive light quarks with
$m_q\sim m_\rho/2$.

Further, along with the production of color-singlet state of $(c\bar q)$-system
we have to take into account the production of color-octet state with the
corresponding transition into the hadron ($D^*$-meson). As we know from the
experience on the consideration of production for the mesons containing two
heavy quarks, the octet term is suppressed by the fourth power of relative
velocity $v$ for the quarks inside the meson \cite{OCT}. In the problem studied
the velocity is not small and the contribution by the octet can be comparable
with the singlet term. Due to the specifics of color amplitude for the
$e^+e^-$-annihilation the ratio between the production of color-singlet and
color-octet $(c\bar q)$-states is fixed, so that introducing the octet term for
the fragmentation we have to substitute for $\langle O_{(\ref{fact})} \rangle$
in
Eqs.(3) and (\ref{prob}) by $\langle O^{eff} \rangle$ as follows:
\begin{equation}
\langle O^{eff}\rangle=\langle O_{(\ref{fact})}\rangle + \frac{1}{8}
\langle O_{(8)} \rangle,
\label{ratio}
\end{equation}
where
$$
\langle O_{(8)} \rangle =
\frac{1}{8 M} \left(-g^{\mu\nu}+\frac{p^\mu p^\nu}{M^2}\right)\;
\langle D^*(p)|(\bar c \gamma_\mu \lambda^a q) (\bar q \gamma_\nu \lambda^b
c)|D^*(p)\rangle\; \frac{\delta^{ab}}{8}.
$$

As we have already mentioned, for the mesons containing the light relativistic
quark we expect $\langle O_{(8)}\rangle / \langle O_{(\ref{fact})} \rangle \sim
1$, since $v\sim 1$. Then from Eq.(\ref{ratio}) we see that at such the ratio
of operators the singlet production dominates in $e^+e^-$-annihilation. The
form of differential cross sections is one and the same for the singlet and
octet mechanisms in the fragmentation model. Therefore, we cannot distinguish
the singlet and octet terms in the production of $D^*$-mesons in the
$e^+e^-$-annihilation. However, in deep inelastic scattering the situation is
essentially changed in the region of $p_T\sim m_c$, where the fragmentation is
not the only mechanism, determining the production of heavy meson.

Thus, we have quite a definite picture for the description of charmed meson
production in perturbative QCD with the factorization of terms responsible for
the contributions by the singlet and octet operators.

Concerning the form of FF in this approach of factorization for the hard
perturbative amplitude and nonperturbative binding operators, we note that the
virtuality of heavy quark is equal to
$$
m_c^2-(P_{D^*}+P_{jet})^2\approx m_{D^*}^2\left (
(1-r)^2-\frac{1}{z}-\frac{r^2}{1-z}\right )= -\frac{m_{D^*}^2}
{z(1-z)}\; (1-(1-r)z)^2,
$$
so that $r^2\approx \epsilon$. At an appropriate choice of $m_q$ and $m_c$ we
can reach the FF behaviour similar with that of FF by Peterson et al. in
(\ref{Peter}). The logarithmic dependence of $\epsilon$ mentioned above can be
reasonably neglected in the region, where $ln \frac{p_T}{m_c}\sim 1$, i.e. it
is not large.

\section{Photoproduction and electroproduction of charm.}

As we have shown above, in the $e^+e^-$-annihilation we obtain the reliable
description of cross sections for the production of charmed particles, if we
use the perturbative calculations with a nonzero mass of light quark. In this
way the calculations are reduced to the evaluation of single diagram in a
special axial gauge (see Fig.~1) \cite{Frag}. In other processes the number of
diagrams is essentially greater than the mentioned one. The complete set of
such diagrams in the $O(\alpha \alpha_s^3)$-order is shown in Fig.~2 for the
subprocess of $\gamma^* g \to D^*+\bar c+q$. Among this set, wherein the
initial photon is on mass-shell for the photoproduction and virtual for the
deep inelastic scattering, we can identify the diagrams corresponding to the
hard production of $c$-quarks up to $O(\alpha \alpha_s)$ with the consequent
fragmentation of heavy quark in the $O(\alpha_s^2)$-order (for example, graphs
16 and 19). These terms dominate at $p_T\gg m_c$, which reproduces the
predictions of fragmentation model by Eq.(\ref{fact}) in the kinematical region
specified.

Let us study the process $\gamma g\to D^*+X$ at a rather high energy 
$\hat s \gg 4 M_{D^*}^2$ to prove the validity of our statement on the
factorization. The large subprocess energy is required to reach the limit of
$p_T \gg m_D$, where the predictions of the model can be compared with the
fragmentation regime in the form of (\ref{fact}).
 
The calculation results for the $p_T$-distribution of the $D^*$-meson
production cross section in the framework of our model in comparison with the
fragmentation prediction (\ref{fact}) are presented in Fig.~3.  In Eq.
(\ref{fact}) the Born approximation for the $c\bar c$-pair production and the
fragmentation function in the form of (3) have been used. We can see that the
factorization for the hard $c\bar c$-pair production and the probability of
$c$-quark fragmentation into $D^*$  for the photoproduction cross section takes
place at $p_T > 20$~GeV for the singlet state and at $p_T>40$~GeV for the octet
term at the photon-gluon energy equal to 200 GeV. We find that at the lower
energy of photon-gluon interaction, say, at $\sqrt{\hat s}=40$ GeV, the
fragmentation regime works at $p_T > 12$ GeV for the singlet state production,
but the differential cross section for the production of octet state cannot be
described by the factorized form of fragmentation since at such the energy the
higher twists over $p_T$ dominate completely in the whole kinematical region.
      
Thus, the model approaches the fragmentation regime at rather large transverse
momenta. At $p_T\sim m_c$ the dominant contribution comes due to the diagrams
of nonfragmentational kind. In these diagrams the independently produced
$c\bar c$ and $q\bar q$ pairs after rescattering are transformed into hadrons.
We call them as the recombination diagrams (see, for example, graphs 5, 6, 7 in
Fig.~2). The fragmentation and recombination diagrams have different forms of
asymptotic behavior at large $p_T$. So, the recombination contribution at
$p_T\gg m_c$ drops like $1/p_T^6$ in contrast to the fragmentation term
decreasing as $1/p_T^4$.      

The fragmentation approach is valid at $p_T\gg m_Q$ and cannot be applied at
$p_T\sim m_Q$ (see Fig.~3). To describe the region of $p_T\sim m_Q$ we have to
take into account subleading terms over $1/p_T$, i.e. the higher twists. The
offered model allows us to get correct factorization at high $p_T$ and to
account for the subleading terms on the basis of gauge invariance dictating the
consideration of complete set of diagrams to the given, fixed, order in
$\alpha_s$.

In the framework of fragmentation approach, the LO calculations must take into
account the resolved photon contribution. In the model under consideration,
which involves the higher orders, this contribution is also included by means
of diagrams, where the splitting of photon into the pairs of light quarks is
taken into account perturbatively at rather high $p_T$, and, hence, at large
virtualities\footnote{We check that the nonperturbative splitting of photon
into the pair of charmed quarks due to the contribution of vector bound states
can be neglected in the kinematical region under consideration.}. 

An analysis of \cite{Frixione} shown that the logarithmic corrections to
the $c$-quark production become essential at $p_T> 20$~GeV. Thus, we can
believe that at moderate $p_T$: $\ln \frac{p_T}{m_c}\le 1$, the fixed order
calculations (LO, NLO) are valid. 

Remember, Eq.(\ref{fact}) cannot be applied at $p_T\sim m_c$ because
of two following reasons:

1) The limit of $p_T\gg m_c$ is not justified at $p_T\sim m_c$,

2) The higher twists over $p_T$ must be taken into account beyond the
fragmentation regime leading to Eq.(\ref{fact}).

In contrast, the offered model under consideration can be used in the kinematic
region $p_T\sim m_c$ too, where Eq.(\ref{fact}) does not work, and it
reproduces (\ref{fact}) at $p_T\gg m_c$.

Thus, exploring the gauge invariance to the fixed order, we extend the
fragmentation term up to the complete set of diagrams to cover the higher
twists over $p_T\sim m_c$.
    
We would like to stress that the perturbative calculations for the production
of $c\bar q$-system can reasonably describe the region of $p_T\sim m_c$, if we
suppose the appropriate effective mass of light quark $m_q$. In the
calculations we chose the value $m_q = 0.3\ {\rm GeV} \sim m_{\rho}/2$. Then
the gluon virtualities are greater than $|k_g^2|\sim p_T^2+m_{\rho}^2$, which
allows us to use pQCD and to stay beyond the region of infrared divergency.

Remember that at $p_T>12$~GeV we normalize the absolute value of cross section
due to relation (\ref{prob}), assuming that the probability of
fragmentation does not depend on the scale.

In contrast to $e^+e^-$-annihilation, in both the photoproduction and the
electroproduction we find another ratio of fragmentational and recombinational
contributions of octet state in comparison with the ratio for the singlet. So,
if $\langle O_{(8)}\rangle / \langle O_{(\ref{fact})} \rangle \sim 1$, then the
recombination contributes to the production of octet with the same order of
magnitude as for the singlet production. Therefore, at the transverse momenta
$p_T\sim m_c$, where the recombination dominates, the values of cross sections
for the singlet and octet are close to each other, when at $p_T\gg m_c$, where
the basic role is played by the fragmentation, the octet state term is
suppressed as $1/8$, according to Eq.(\ref{ratio}).

We emphasize that the introduction of octet term does not practically change
the cross section for the production of $D^*$-mesons at $p_T\gg  m_c$, however,
it essentially contributes in the region of $p_T\simeq m_c$.

\subsection{The photoproduction of $D^*$-mesons.}

We fix the value of fragmentation probability $w(c\to D^*)$ taken from the data
on the $D^*$-meson production in $e^+e^-$-annihilation at LEP, and we put the
choice of $\mu_R$, $m_q$, $m_c$, so that Eq.(\ref{prob}) definitely determines
the value of $\langle O^{eff}(\mu_R)\rangle$ operator matrix element. So, if
\begin{equation}
\begin{array}{rcl}
\mu_R &=& m_{D^*},\\
m_q &=& 0.3 \ {\rm GeV},\\
m_c &=& 1.5 \ {\rm GeV},\\
w(c\to D^*) &=& 0.22.\\
\end{array}
\end{equation}
then
$$\langle O^{eff}(m_{D^*}) \rangle  = 0.25 \ {\rm GeV}^{3}.$$

Such the choice has allowed us to describe the data on the photoproduction of
$D^*$-mesons. The results of calculations are shown in Fig.~4 in comparison
with the ZEUS data \cite{ZEUS} measured in the following kinematical region:
$p_T>2$ GeV, $-1.5<\eta<1.5$, $130\ {\rm GeV}<W<280\ {\rm GeV}$,
$Q^2<1\ {\rm GeV^2}$, where $W$ is the total energy of $\gamma p$-interactions,
$Q^2$ is the photon virtuality, and $\eta$ is the pseudorapidity of
$D^*$-meson.

We see in Fig.~4 that the value of production cross section for the
$D^*$-mesons cannot be explained by the singlet term only. The relative value
of the octet contribution $\langle O_{(8)} \rangle / \langle O_{(\ref{fact})}
\rangle =1.3$ allows us to reach the good agreement with the experimental
spectra of $D^*$-mesons. We put the attention to the mentioned distinct
difference in the behaviour of distributions over the transverse momentum for
the singlet and octet terms. The contribution of octet state is enforced in
the forward direction ($\eta > 0$). At the chosen values of operator matrix
elements the contributions of singlet and octet are close to each other at
$p_T\sim m_c$, while at $p_T\gg m_c$ the octet term becomes less essential than
the singlet one.

We can see that the accounting of  the octet contribution essentially improves
the form of predicted cross section distribution over the pseudorapidity in the
forward direction. Such the improvement is due to the fact that the
recombination diagrams dominant at $p_T\sim m_c$ are enforced in the octet part
of the cross section by the color factor.

The interaction of virtual $c$-quark with the light valence quark in the
initial hadron can be taken into account easily, as we will show below, but
this process becomes essential only at low energies of charm production near
the threshold.    

The scale of factorization for the gluon structure function in the proton has
been fixed $\mu_F = 2m_{D^*}$. Calculating the differential cross sections we
suppose two values for the scale of factorizing the matrix elements for the
quark operators $\langle O_{(1,8)} \rangle  $:  $\mu_R = m_{D^*}$ (the upper
line) and $\mu_R = 2m_{D^*}$ (the down curve). We see that the experimental
data on the photoproduction of $D^*$-mesons at HERA can be described in the
model in the better way if we put $\mu_F =2m_{D^*}$ and $\mu_R = m_{D^*}$.

The predictions for the photoproduction of $D^*$-mesons in the low energy
region studied by ZEUS \cite{ZEUSN}: $p_T>2$ GeV, $-1.<\eta<1.5$, $80\ {\rm
GeV}<W<120\ {\rm GeV}$, $Q^2<0.01\ {\rm GeV^2}$, are shown in Fig.~5. In the
calculations we use the same values of $\langle O_{(\ref{fact})} \rangle$ and
$\langle O_{(8)} \rangle$ as they stand above.

Unfortunately, the present experimental data do not yet allow us to extract the
value of $\langle O_{(8)} \rangle$ with a low uncertainty, and with the
mentioned variety in the factorization scales we can write down \cite{p1}
$$
\langle O_{(8)}\rangle \approx 0.33-0.49 \ {\rm GeV^3.}
$$
It is quite clear that changing the parameter $\langle O_{(\ref{fact})}
\rangle/\langle O_{(8)} \rangle$ we can enhance or decrease the contribution of
higher twists in the region of $p_T\sim m_c$. In the calculations we neglect
the difference between the spectra of vector and pseudoscalar octet states,
whose contributions are effectively summed up in the corresponding operator.
Furthermore, we do not consider a possible influence of $P$-wave color-octet
states, since their contribution is suppressed in comparison with the $S$-wave
contributions, as we saw in similar studies \cite{ZP}.

\subsection{The photoproduction of $D_s$- and $D_s^*$-mesons.}
 
In the model under consideration we predict the cross section for the
photoproduction of $D_s$- and $D_s^*$-meson. The general features of these
processes were discussed in our work \cite{pBKL}. A more detailed analysis
is given below.

The probability of $c$-quark fragmentation into $D_s$ or $D_s^*$,
$w_s(c\to D_s, D_s^*)=0.1$, was fixed from the data on the $D_s$-meson
production in the $e^+e^-$-annihilation at LEP \cite{LEPDs}.   

Using this value of $w_s$ and replacing the effective valence quark mass
$m_q=0.3$~GeV by $m_s=0.5$~GeV in the amplitude of $D^*$-production, we
calculate the expectation of differential cross section for $D_s^*$. The
analogous technique was used for the calculation of $D_s$ production cross
section. In the both cases the value 1.3 was chosen for the relative octet
contribution such as in  the $D^*$-meson production. It is worth to mention
that the flavor $SU(3)$-symmetry for the $u,d,s$ quarks is broken. So, to
obtain the experimentally measured fraction of charmed-stranged events in the
charm production, $w_s$, we put the effective operator $\langle
O^{eff}_s(m_{D^*_s})\rangle$ for the $c$-quark fragmentation into $D_s$ to be
approximately twice greater than $\langle O^{eff}(m_{D^*})\rangle$.  
 
In this work we present the differential cross section for the sum of $D_s$-
and $D_s^*$-mesons because the experimental discrimination between the events
with $D_s$ and $D_s^*$ is problematic. The reason is the necessary registration
of a soft photon coming from the dominant mode of $D_s\to D_s^*\gamma$ decay,
which is quite difficult task.

The results of our calculations in comparison with the ZEUS experimental data
\cite{ZEUSDs} are presented in Fig.~6 for the following kinematical region:
$130<W<280$~GeV, $Q^2<1\ {\rm GeV}^2$, $p_T>3$~GeV and $-1.5<\eta<1.5$.
 
We see that the predictions for the $D_s$ and $D^*_s$ photoproduction
cross section are in a good agreement with the experiment.
 
\subsection{Electroproduction of $D^*$-mesons.}

In contrast to the photoproduction, we have to take into account the exact
$Q^2$ dependence in DIS for the calculation of $D^*$ production. In this
respect we evaluate the tensor of leptonic currents squared and averaged over
the polarizations to contract it with the photon-gluon vertices in the charm
production calculated numerically at arbitrary $Q^2$ of photon. Again we
consider the production of $c\bar q$-pair in the $O(\alpha_{em}^2
\alpha_s^3)$-order.

The used technique of calculations allows us to evaluate the cross sections for
the production of $D^*$-mesons in the case when the initial photon is
essentially virtual, i.e. off mass shell. For this purpose the matrix element
used in the photoproduction $M_i$ taken at the photon virtuality $Q^2$ not
equal to zero, where $i$ is the index corresponding to the photon polarization,
is squared and convoluted with the spin averaged electron current squared in
the following way:
\begin{equation}
|A|^2=\sum_{ij}
\frac{k_1^i k_2^j+k_1^jk_2^i-\frac{Q^2}{2}g^{ij}}
{Q^4}M_i(Q^2) M_j^*(Q^2) ,
\end{equation}
where $|A|^2$ is the square of amplitude for the electroproduction of
$D^*$-mesons, $k_1$ and $k_2$ are the initial and final momenta of positron,
and $Q^2=-(k_1-k_2)^2$.

The results of calculations are presented in Fig.~7 for the production of
$D^*$-mesons in deep inelastic scattering of $e^+p$ in comparison with the
experimental data of ZEUS Collaboration \cite{DIS} in the following kinematical
region: $1\ {\rm GeV}^2 < Q^2 < 600\ {\rm GeV}^2$, $-1.5 <\eta < 1.5$,
$1.5 \  {\rm  GeV}  <  p_T  <  15\ {\rm GeV}$,  $0.02 < y < 0.7$,  where
$y=W^2/s$.

Calculating the matrix element we use the running coupling constant of QCD at
the scales $\mu_R = \sqrt{m_{D^*}^2+Q^2}$ (the upper curve) and $\mu_R =
\sqrt{4m_{D^*}^2+Q^2}$ (the lower curve), and the structure function of gluon
inside the proton is evaluated  at the scale $\mu_F = \sqrt{4m_{D^*}^2+Q^2}$ in
the CTEQ4 parameterization \cite{CTEQ}. Such the choice allows us, first,
to get a good description of electroproduction and, second, to reach the scales
used in the photoproduction at $Q^2=0$.

The ratio of $\langle O_{(\ref{fact})} \rangle/\langle O_{(8)} \rangle =1.3$
has been supposed the same as in the photoproduction in the good agreement with
the data.

\section{The total cross section of charm photoproduction.}

Since we see quite a good description of current data on the photoproduction of
charm at $p_T\sim m_c$ in the model, it would be of interest to test its
predictions for the total cross sections of charm photoproduction in a broad
region of interaction energy \cite{LOW}. We evaluate the total cross sections
of charmed particles at the $\gamma p$-energies in the interval from 20 to 300
GeV (see Fig.~8).

To estimate the total charm yield we multiply the $D^{*+}$ and $D^{*-}$ cross
section by the factor of $1/0.44$ being the inverse fraction of these particles
in the charm events as was measured in the $e^+e^-$annihilation. Furthermore,
to compare with the data at low energies we have to take into account the
additional contribution of diagrams related to the production of $D^*$-mesons
off the valence quarks in the subprocess $\gamma  q_v\to (\bar c q) +c\to
D^{*-}+X$. The consideration of such term can be performed with no introduction
of additional parameters because the normalization of the cross section is
determined by the same factor of $\langle O^{eff}\rangle$. In contrast to the
contribution of $\gamma g \to (\bar c q) +c+\bar q$ subprocess, the
contribution of $\gamma  q_v \to (\bar c q) +c$ weakly depends on the light
quark mass $m_q$. However, due to the additional condition (\ref{prob})
relating the values of $m_q$ and $\langle O^{eff}\rangle$ the implicit
dependence of charm yield on $m_q$ takes place for the production on the
valence quark. So, changing the light quark mass from 0.3 to 0.26~GeV  the
contribution of this subprocess decreases approximately by 1.5 times. The other
terms remain unchanged in this way.

Thus, varying the mass $m_q$ we can adjust the contribution of the $\gamma
q_v\to (\bar c q) +c\to D^{*-}+X$ subprocess due to the normalization factor of 
$\langle O^{eff} \rangle$, and the slow overestimate of experimental data
can be removed by the small correction of $m_q$ value.

At the HERA energies the contribution of valence quarks is about several
percents, and in the framework of present paper we consider a further
discussion on this question is inappropriate.

\section{Conclusion.}

We have presented the model for the production of charmed mesons in the
positron-proton scattering. The essential ingredient of consideration is the
introduction of fast degrees of freedom inside the heavy meson for both the
heavy and light quarks, whose differential characteristics are calculated in
the framework of perturbative QCD at the transverse momenta $p_T\ge m_c$. In
the model the hadronization is described by means of matrix elements in the
form of factorizing the hard partonic subprocess and the nonperturbative quark
operators corresponding to the transition of quark state into the hadron. In
this approach, the semilocal duality assumes that the differential
characteristics of quark pair composing the meson repeat the distributions of
hadron in the region of applicability of perturbative QCD $p_T\ge m_c$. At
large $p_T\gg m_c$, as we expect from the common theorem on the factorization
of hard processes, the cross section has the form of convolution for the cross
section of heavy quark production with the fragmentation function.

The region of $p_T \sim m_c$ is described by additional terms different from
the fragmentation. They have the asymptotics $1/p_T^6$. We enhance the charm
yield in this region due to the contribution coming from the color-octet $c\bar
q$-state.

Within the model uncertainties we have described the data on the
photoproduction and electroproduction of $D^*$-mesons at the HERA energies as
well as the data on the total cross sections of charm photoproduction at fixed
target.

The improvement of $D^*$-meson spectrum description in the present model in
comparison with other approaches is achieved due to the clear physical effect:
the interaction of virtual $c$-quark with the fast light quark is increased in
the color-octet state by the diagrams of recombination in the charm production
at $p_T\sim m_c$. At large $p_T$ this contribution becomes negligible  and the
fragmentation regime dominates. It is worth to mention that in our approach the
calculation of light quark distribution at large $p_T\sim m_c$ has been done
perturbatively. If one neglects the interference terms, then this mechanism for
the enhancement of charm yield in the forward direction can be interpreted as
the $c$-quark interaction with the hadronic remainder (the sea of light
quarks), which increase the forward production (so-called "drag-effect")
\cite{Harris}.     

The most essential and main advantage of the method used is the possibility to
take into account the interferences of different contributions as well as the
higher twists over the transverse momentum.

We insist on the use of `fragmentation' term as not equivalent to
`hadronization' in a general case. For example, Fig.~5 shows the most
successful theoretical description of rapidity distributions for the
photoproduction of $D^*$, and the deviations at large rapidities are not so
dramatic as in other calculations cited above. They explored a point of view
coming from a believement: the hadronization implies the fragmentation. In
order to describe the fragmentation of heavy quark, nothing is required beyond
the perturbative QCD cross section of single heavy quark convoluted with a
well-justified function by Peterson et al., say. To consider the general
hadronization of fast components in the heavy meson, one should introduce some
physical quantities defining the fastness (cut off low virtualities, or the
constituent mass of light quark) and the general probabilities to find the fast
components inside the meson in the color singlet and octet states. So, the
model parameters are degenerate into the minimal number (3) of unavoidable
quantities in the hadronization of heavy meson. Moreover, the data on the
electon-positron annihilation into $D^*$ fix the sum of octet and singlet
terms. The normalization of cross section for the photoproduction of $D^*$ at
large transverse momenta and high energies of photon-nucleon collisions
restricts the variation of constituent mass, while the $p_T$-distribution
measures the normalization of octet contribution. Then we reproduce the
differential cross sections on the rapidity in various ranges of transverse
momenta (3 distributions) and photon-nucleon collision energies as well as in
DIS (6 distributions) with a good accuracy. This fact implies the success of
the model in contrast to other QCD-inspired approaches based on the
fragmentation: they use the variation of charmed quark mass to get a good
normalization with large diviations in distributions or fit the high $p_T$
distribution with a loss of correct noramalization of integrated cross section.

It is quite evident that the offered scheme of evaluation for the additional
contribution into the production of heavy mesons can be transformed to the case
of hadronic production. In this sense, it would be of interest to compare such
estimates with the data on the $B$-meson yield at the FNAL Tevatron, which we
plan to make in the forthcoming papers.

This work was in part supported by the Russian Foundation for Basic Research,
grants 99-02-16558 and 00-15-96645.

The authors express their gratitude to L.K.Gladilin (ZEUS) for the help in
the investigations and the fruitful discussions, to S.S.Gershtein and G.Kramer
for the discussions and useful advices.

\small
\newpage
\includegraphics{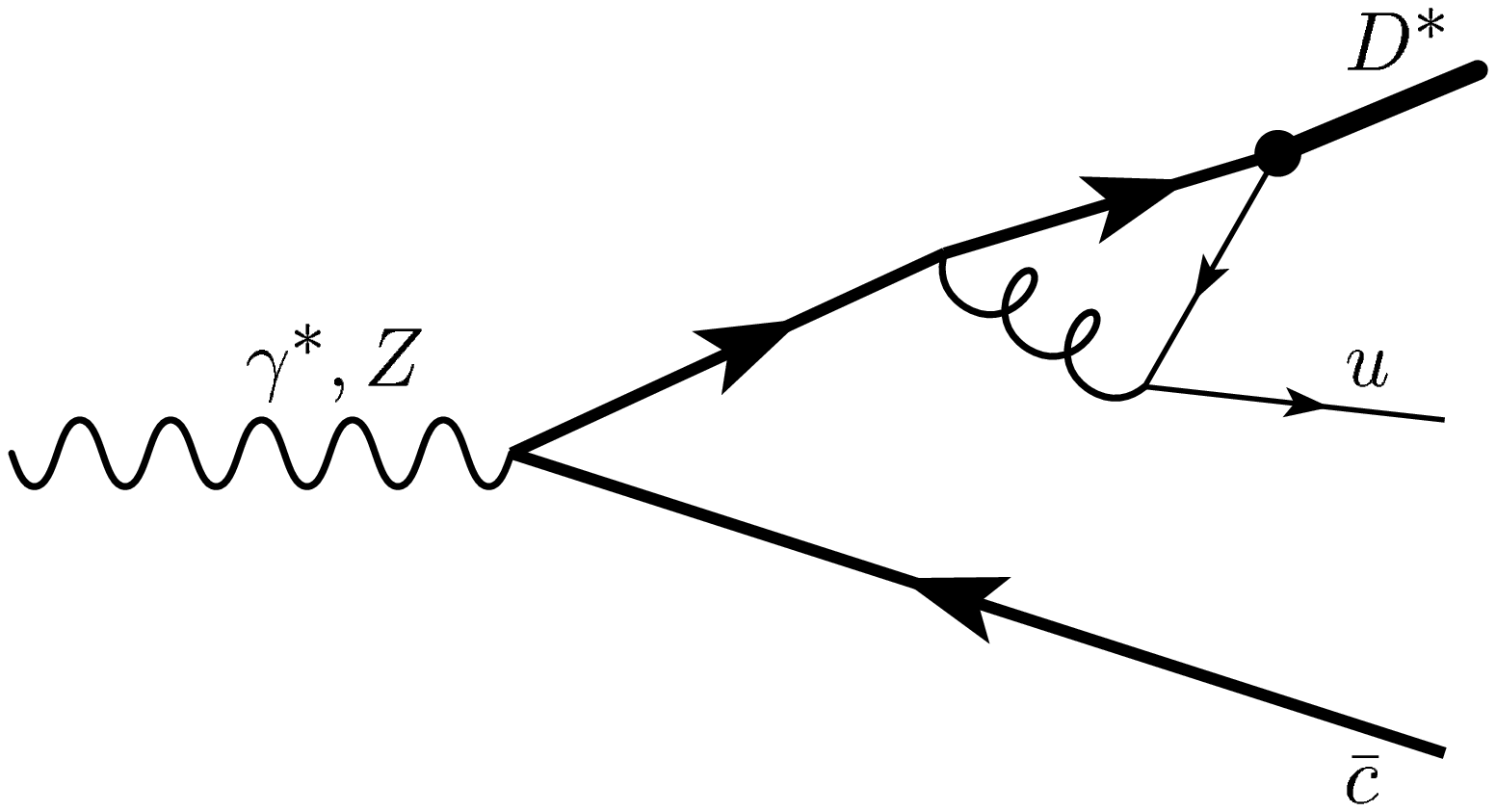}
\vspace*{12 cm}
\hspace*{1 cm} Fig.~1. \parbox[t]{12cm}{The diagram describing the production
of $D^*$-meson in the process of $e^+e^-\to \gamma^*,Z\to D^*+X$ to the leading
order.}

\newpage
\includegraphics{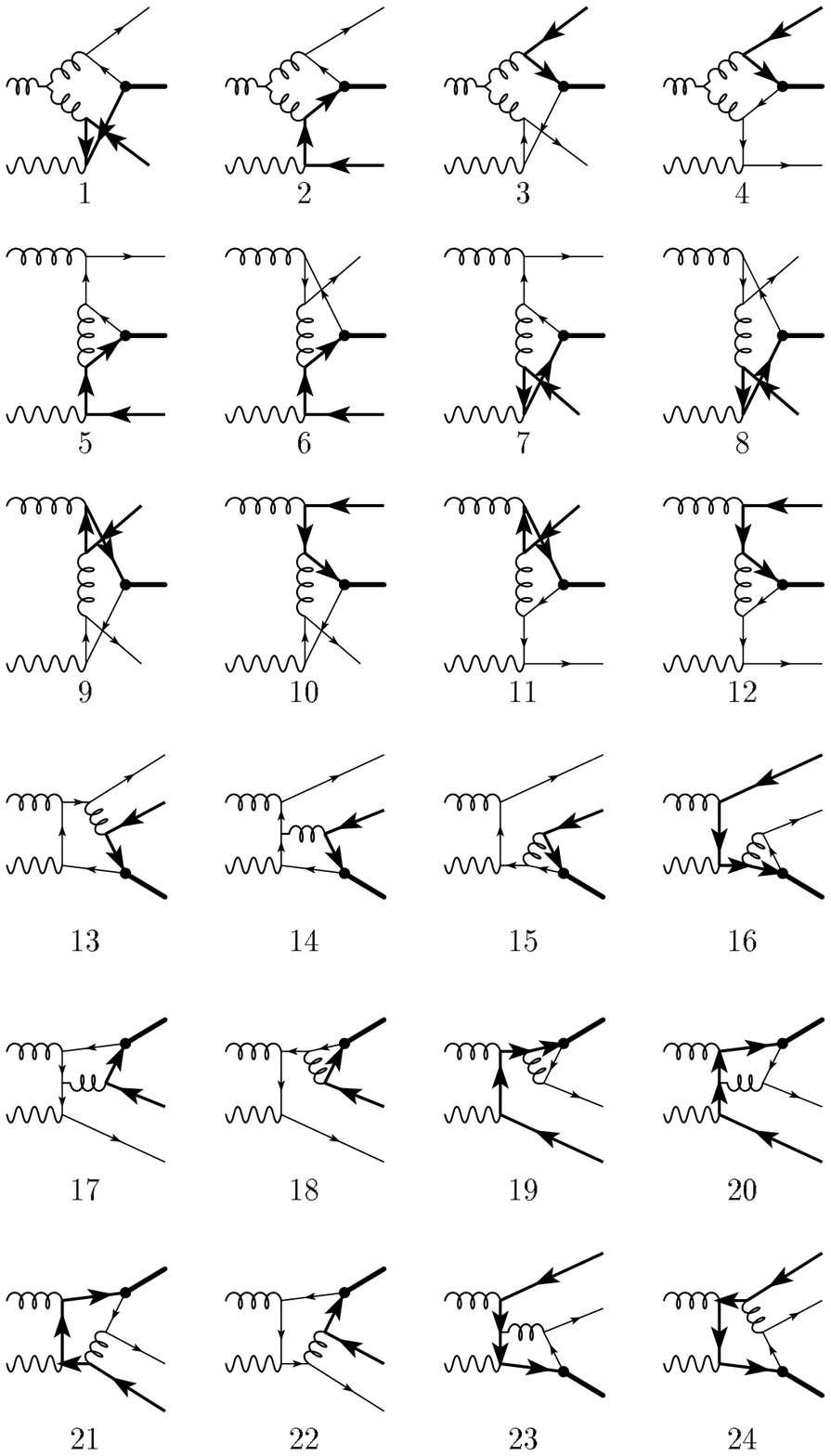}
\vspace*{21 cm}
\hspace*{2 cm} Fig.~2. \parbox[t]{12cm}{The leading order diagrams for the
production of $(c\bar q)$-state in the $g \gamma^*$-interactions.}

\newpage
\vspace*{-3cm}
\includegraphics{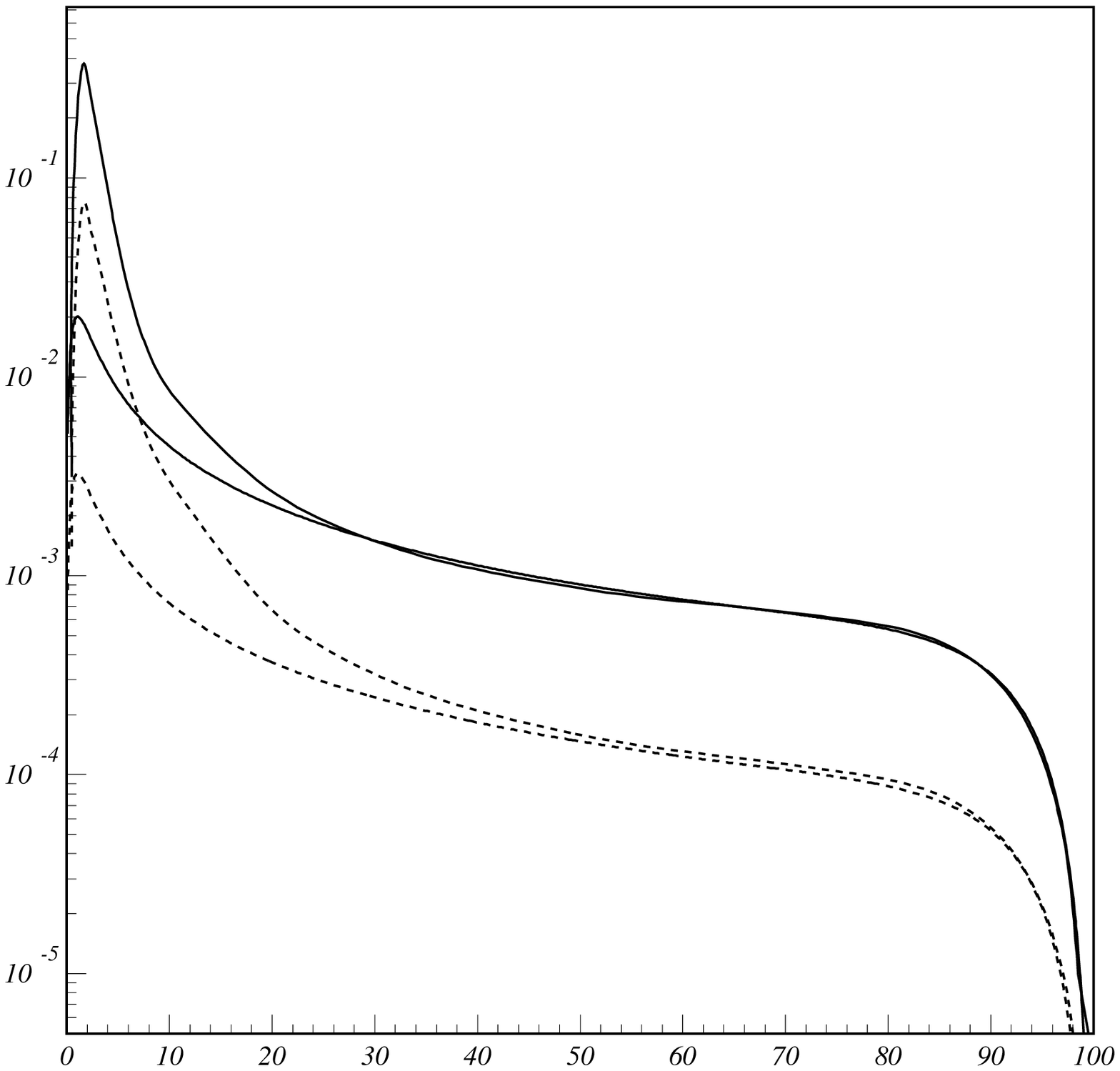}

\begin{picture}(450,450)
\put(0,320){$d\sigma_{g\gamma}/dp_T$, nb/GeV}
\put(340,-50){GeV}
\put(20,-80)
{Fig.~3. \parbox[t]{12cm}{The distribution over the transverse momentum of
$D^*$-mesons produced in the singlet $c\bar q$-state (the upper solid curve) in
comparison with the fragmentation model (the down solid curve) in the $\gamma
g$-interaction at the energy 200 GeV. The analogous distributions are shown for
the $D^*$-meson production in the octet $c\bar q$-state (the dashed curves). }}
\end{picture}

\newpage
\vspace*{-3 cm}
\includegraphics{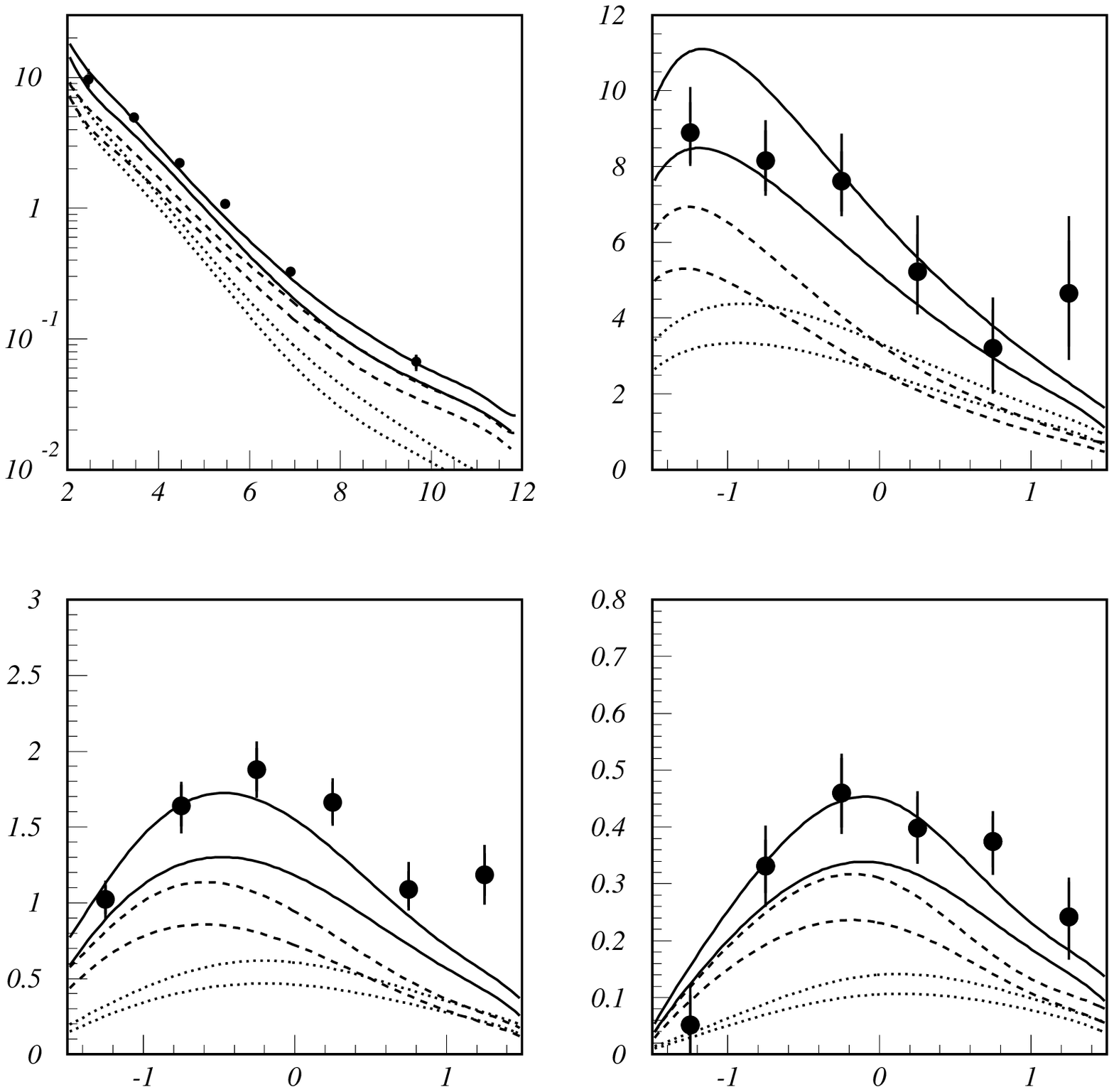}

\begin{picture}(450,450)
\put(-10,350){$d\sigma/dp_T$, nb/GeV}
\put(240,350){$d\sigma/d\eta$, nb}
\put(-10,100){$d\sigma/d\eta$, nb}
\put(240,100){$d\sigma/d\eta$, nb}
\put(150,120){$p_T$, GeV}
\put(400,120){$\eta$}
\put(150,-130){$\eta$}
\put(400,-130){$\eta$}
\put(345,300){$p_T>2$ GeV}
\put(345,50){$p_T>6$ GeV}
\put(95,50){$p_T>4$ GeV}
\put(10,-180){Fig.~4. \parbox[t]{14cm}{
The differential distributions in the photoproduction of $D^*$-meson over the
transverse momentum ($p_T$) and the pseudorapidity ($\eta$) in comparison with
the ZEUS data at $130<W<280 \ {\rm GeV}$ and $Q^2<1\ {\rm GeV^2}$. 
The dashed line presents the contribution of the color-singlet term, the dotted
line does the octet term, and the sum of terms is shown by the solid line. The
upper and down curves correspond to two scales for the factorizing the matrix
elements of quark operators as it is described in the text.
}}
\end{picture}

\newpage
\vspace*{-3cm}
\includegraphics{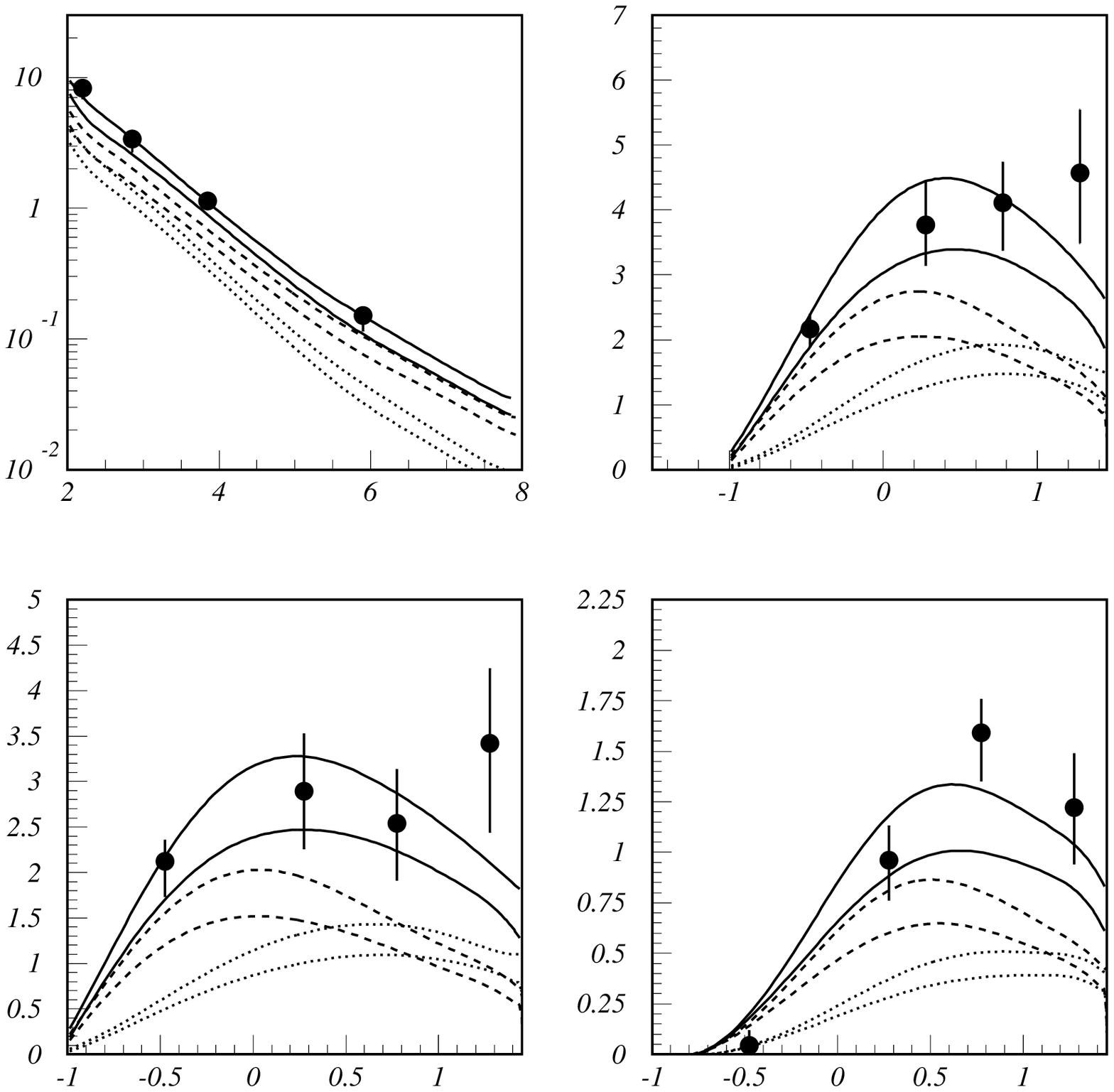}

\begin{picture}(450,450)
\put(-10,350){$d\sigma/dp_T$, nb/GeV}
\put(240,350){$d\sigma/d\eta$, nb}
\put(-10,100){$d\sigma/d\eta$, nb}
\put(240,100){$d\sigma/d\eta$, nb}
\put(150,120){$p_T$, GeV}
\put(400,120){$\eta$}
\put(150,-130){$\eta$}
\put(400,-130){$\eta$}
\put(255,310){$2<p_T<8$ GeV}
\put(10,60){$2<p_T<3.25$ GeV}
\put(255,60){$3.25<p_T<8$ GeV}
\put(20,-180){Fig.~5. \parbox[t]{12cm}{
The differential distributions in the photoproduction of $D^*$-mesons at $80
<W<120 \ {\rm GeV}$ and $Q^2<0.01\ {\rm GeV^2}$. The notations are the
same as in Fig.~4.
}}
\end{picture}

\newpage
\vspace*{-3 cm}
\includegraphics{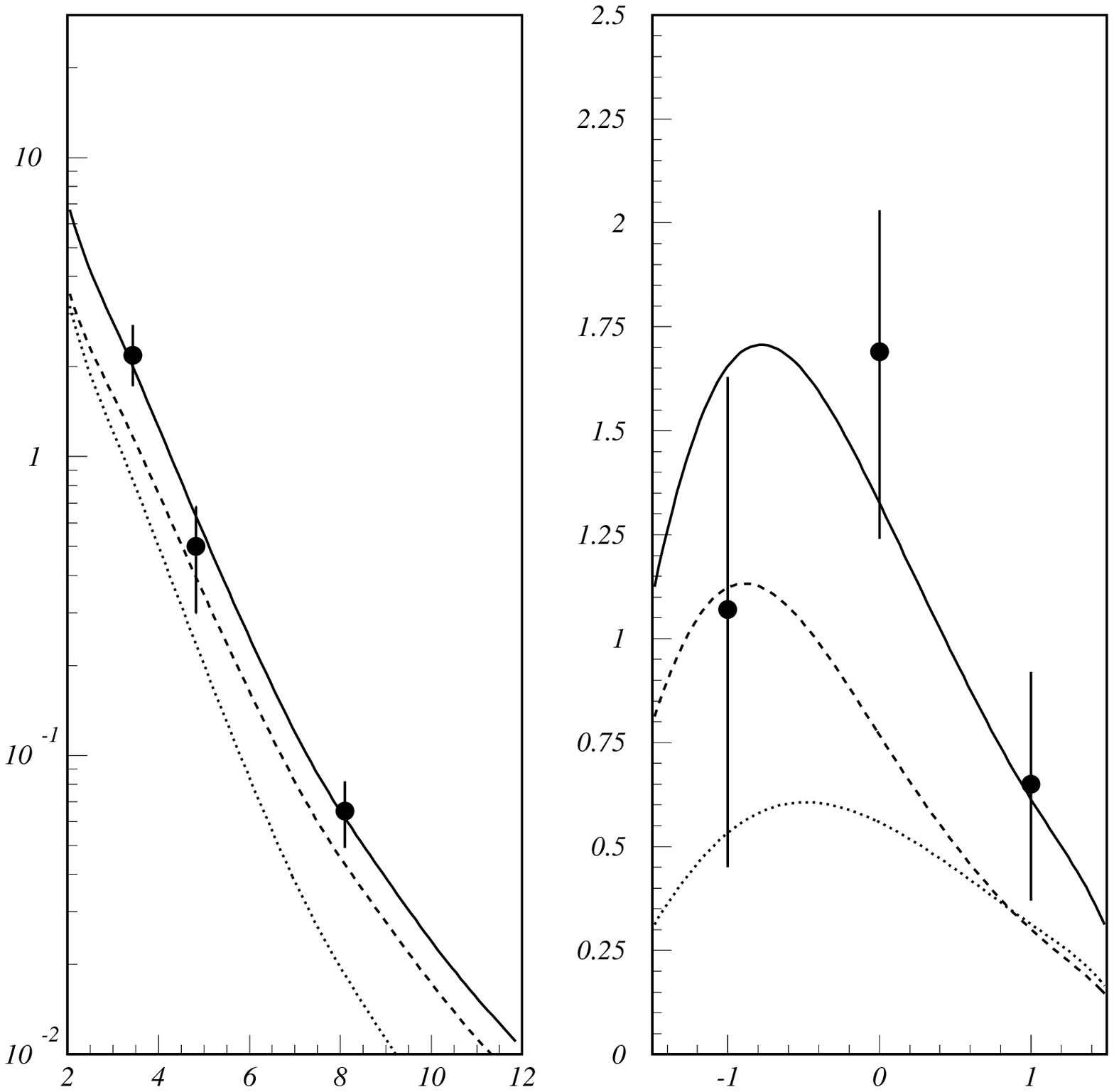}

\begin{picture}(450,450)
\put(-10,250){$d\sigma/dp_T$, nb/GeV}
\put(240,250){$d\sigma/d\eta$, nb}
\put(150,-30){$p_T$, GeV}
\put(400,-30){$\eta$}
\put(345,180){$p_T>3$ GeV}
\put(20,-180){Fig.~6. \parbox[t]{12cm}{
The differential distributions in the photoproduction of $D_s$ and
$D_s^*$-mesons at $80<W<120 \ {\rm GeV}$ and $Q^2<0.015\ {\rm GeV^2}$.
The notations are the same as in Fig.~4.
}}
\end{picture}

\newpage
\vspace*{-3cm}
\includegraphics{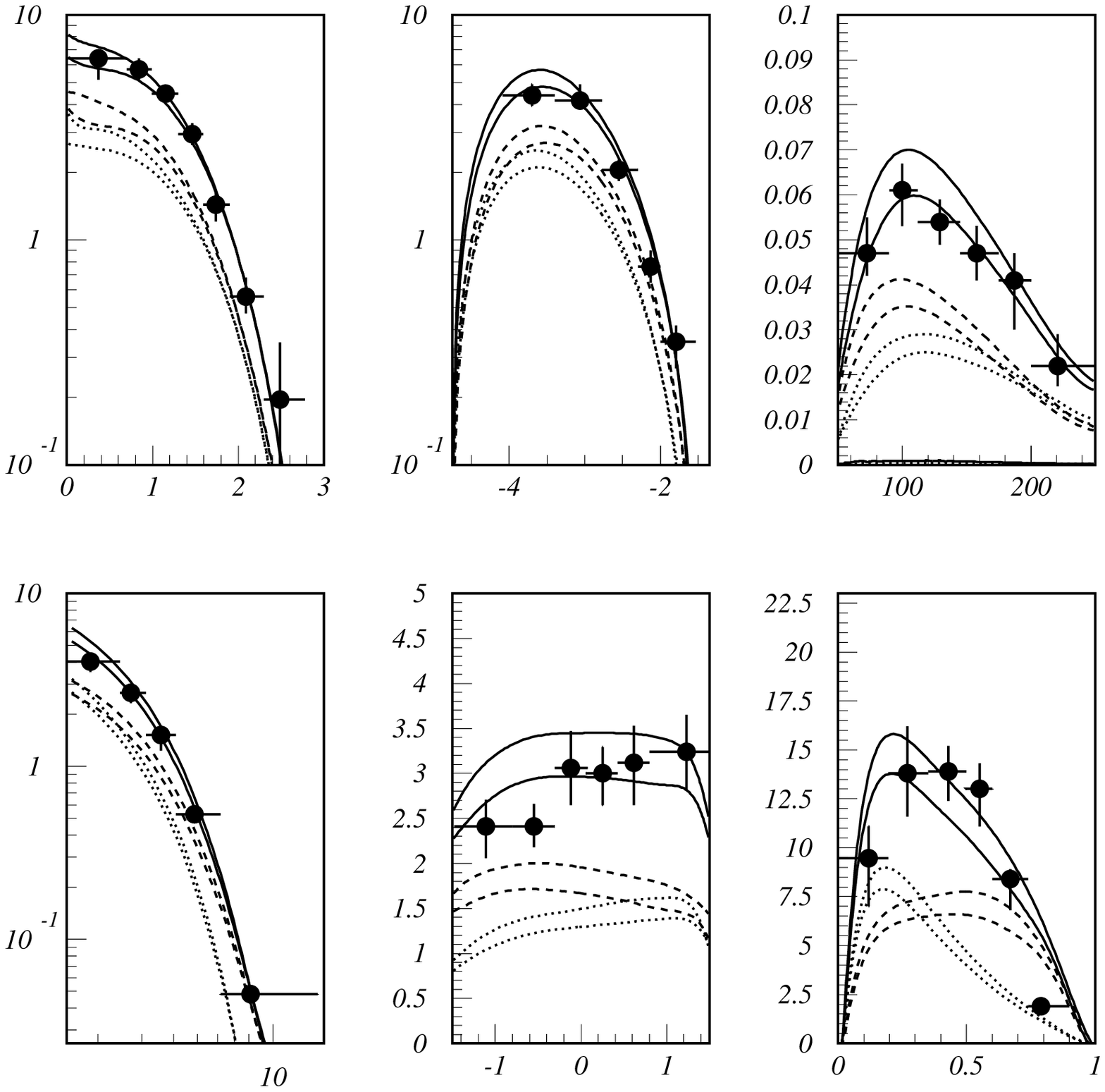}

\begin{picture}(450,450)
\put(-30,350){$d\sigma/dlog_{10}Q^2$, nb}
\put(140,350){$d\sigma/dlog_{10}x$, nb}
\put(305,350){$d\sigma/dW$, nb/GeV}
\put(-30,100){$d\sigma/dp_T$, nb/GeV}
\put(140,100){$d\sigma/d\eta$, nb}
\put(305,100){$d\sigma/dx(D^*)$, nb}
\put(70,120){$log_{10}Q^2$}
\put(240,120){$log_{10}x$}
\put(405,120){$W$, GeV}
\put(70,-130){$p_T$, GeV}
\put(240,-130){$\eta(D^*)$}
\put(405,-130){$x(D^*)$}
\put(70,300){\large \it a}
\put(240,300){\large \it b}
\put(405,300){\large \it c}
\put(70,50){\large \it d}
\put(240,50){\large \it e}
\put(405,50){\large \it f}
\put(20,-180){Fig.~7. \parbox[t]{12cm}
{The differential cross sections for the production of $D^*$-mesons ({\it a} --
over the photon virtuality in GeV$^2$, {\it b} -- over the Bjorken $x$, {\it
c} -- over the invariant mass of hadrons in the final state, {\it d} -- over
the transverse momentum, {\it e} -- over the pseudorapidity, {\it f} -- over
the Feynman $x$) in the deep inelastic $e^+p$-scattering in comparison with the
data of ZEUS Collaboration. The notations are the same as in Fig.~4.}}
\end{picture}

\newpage
\vspace*{-3cm}
\includegraphics{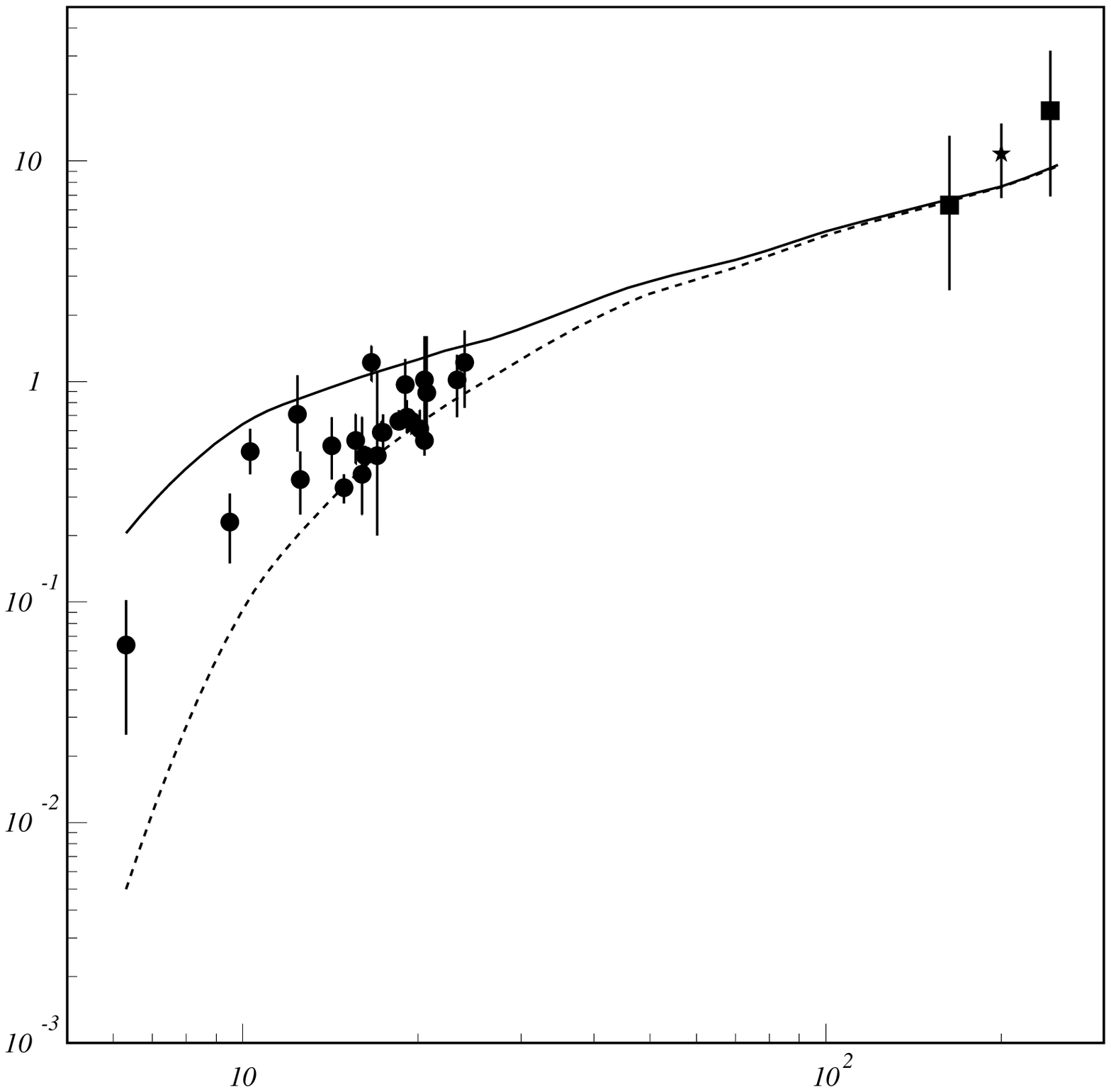}

\begin{picture}(450,450)
\put(0,250){$\sigma_{\gamma p}$, $\mu$bn}
\put(330,-130){GeV}
\put(20,-180)
{Fig.~8. \parbox[t]{12cm}{The total cross section of charm photoproduction
versus the energy of $\gamma p$-interaction $W$. The solid markers represent
the data, and the dashed and solid curves show the model predictions with no
valence contribution and its addition, respectively. }}
\end{picture}

\end{document}